\documentclass[%
reprint,
superscriptaddress,
showpacs,
 amsmath,amssymb,
 aps,
pre,
article
]{revtex4-1}

\usepackage{graphicx}
\usepackage{dcolumn}
\usepackage{bm}

\usepackage{comment} 
\usepackage{color}

\begin{document}

\preprint{APS/123-QED}

\title{Detachment, Futile Cycling and Nucleotide Pocket Collapse in Myosin-V Stepping}

\author{Neville J. Boon}
\email{n.boon@imperial.ac.uk}
\affiliation{
 Department of Mathematics, University of Surrey, Guildford, Surrey GU2 7XH, UK.
}
\affiliation{
 Department of Bioengineering, Imperial College London, London SW7 2AZ, UK.
}

 \author{Rebecca B. Hoyle}
 \email{R.B.Hoyle@soton.ac.uk}
\affiliation{
 Department of Mathematics, University of Surrey, Guildford, Surrey GU2 7XH, UK.
}
\affiliation{
 Mathematical Sciences, University of Southampton, Highfield, Southampton SO17 1BJ, UK.
}

\date{\today}

\begin{abstract}
Myosin-V is a highly processive dimeric protein that walks with 36nm steps along actin tracks, powered by coordinated ATP hydrolysis reactions in the two myosin heads. No previous theoretical models of the myosin-V walk reproduce all the observed trends of velocity and run-length with [ADP], [ATP] and external forcing. In particular, a result that has eluded all theoretical studies based upon rigorous physical chemistry is that run length decreases with both increasing [ADP] and [ATP].

We systematically analyse which mechanisms in existing models reproduce which experimental trends and use this information to guide the development of models that can reproduce them all. We formulate models as reaction networks between distinct mechanochemical states with energetically determined transition rates. For each network architecture, we compare predictions for velocity and run length to a subset of experimentally measured values, and fit unknown parameters using a bespoke MCSA optimization routine. Finally we determine which experimental trends are replicated by the best-fit model for each architecture. Only two models capture them all: one involving [ADP]-dependent mechanical detachment, and another including [ADP]-dependent futile cycling and nucleotide pocket collapse. Comparing model-predicted and experimentally observed kinetic transition rates favors the latter.
\end{abstract}

\maketitle

\section{Introduction}

Gene transcription, directional intracellular transport and cell division are examples of important molecular processes required by all living organisms and performed by motor proteins at a molecular level through the transformation of chemical energy from ATP hydrolysis into mechanical work. Myosin-V is one such motor that walks hand-over-hand along an actin filament taking steps of 36nm \cite{Mehta1999,mehta2001,Warshaw2005}. The two heads of the protein attach and detach from the track in a mechanochemically-coordinated manner to ensure both motion towards the barbed (or plus) end of the actin and that many successive steps are taken before detachment \cite{rosenfeld2004,purcell2005,Oguchi:2008we,Sakamoto2008}.

Over the last two decades, experimental studies have focused upon characterizing the behavior of myosin-V through dynamical walking experiments \cite{forkey2003,Baker2004,uemura2004,veigel2005,clemen2005,Gebhardt:2006tu,Kad:2008wh,Kad:2012ki}, kinetic experiments \cite{DeLaCruz1999,trybus1999,DeLaCruz2000,wang2000,rosenfeld2004,yengo2004} and other measures of stepping mechanics \cite{veigel2002,rosenfeld2004,purcell2005,cappello2007,Oguchi:2008we,Kodera:2010cj}. However, this work has not yet fully unified our understanding of the underlying physical chemistry with all the experimentally observed behavior.

Many models of myosin-V stepping exist within the literature \cite{Fisher1999,rief2000,Kolomeisky2003,rosenfeld2004,Baker2004,vilfan2005,Skau2006,Gebhardt:2006tu,tsygankov2007,Kolomeisky:2007ff,wu2007,Vilfan2009,Bierbaum:2011dj,Kad:2012ki,craig2009, Vilfan2009,Hinczewski2013}, but to the best of our knowledge a satisfactory biomechanochemical description that qualitatively matches all available dynamical data has not yet been proposed. Explaining the experimentally observed average run length before detachment \cite{Baker2004,Kad:2012ki} against both [ADP] \cite{Skau2006} and [ATP] \cite{Bierbaum:2011dj,Baker2004,wu2007} simultaneously has proved a considerable challenge. Furthermore, many models do not explicitly account for the underlying physical chemistry that places important restrictions on rate constants which can have a large effect on the described behavior.

In this article we compare existing myosin-V models within a single mathematical framework for the first time. Comparisons are performed between models using the same set of differential equations to describe each model, with appropriate choices of parameters in each case. This allows us to ascertain which mechanisms included in existing models lead to reproduction of which experimental trends and hence to guide development of models that can reproduce them all. We use optimization techniques to match a model of a given architecture as closely as possible to experimental data, and this reveals that certain architectures or combinations of mechanisms simply cannot give rise to certain experimental trends. We emphasize that this is not simply an exercise in parameter-fitting, but rather a systematic and informed exploration of model-space that allows us to unpick and rebuild model architectures - in terms of their reaction pathways - in order to match the available experimental observations. In this way we deduce energetic descriptions of myosin-V stepping that comprehensively capture the motor's qualitative dynamical behavior for the first time. 

\section{Model space}

Our approach to model development has three aspects. Firstly the identification of an appropriate model space. Secondly an optimisation routine that identifies the closest match for a given point in model space to a subset of the available experimental data. Finally, a comparison between the qualitative behaviour of the full set of experimental data and the optimised model to indicate the direction in model space in which we should move.

We begin by describing the mechanisms that we include in candidate model architectures and use these to explore model space revealing two possible descriptions that achieve our aim. Comparing the kinetic transition rates predicted by these two models with experimentally observed values lends tentative support to a mechanism including nucleotide-dependent futile cycling with nucleotide pocket collapse over one that involves mechanical motor detachment.
 
\begin{figure}
\begin{center}
\includegraphics[width=3in]{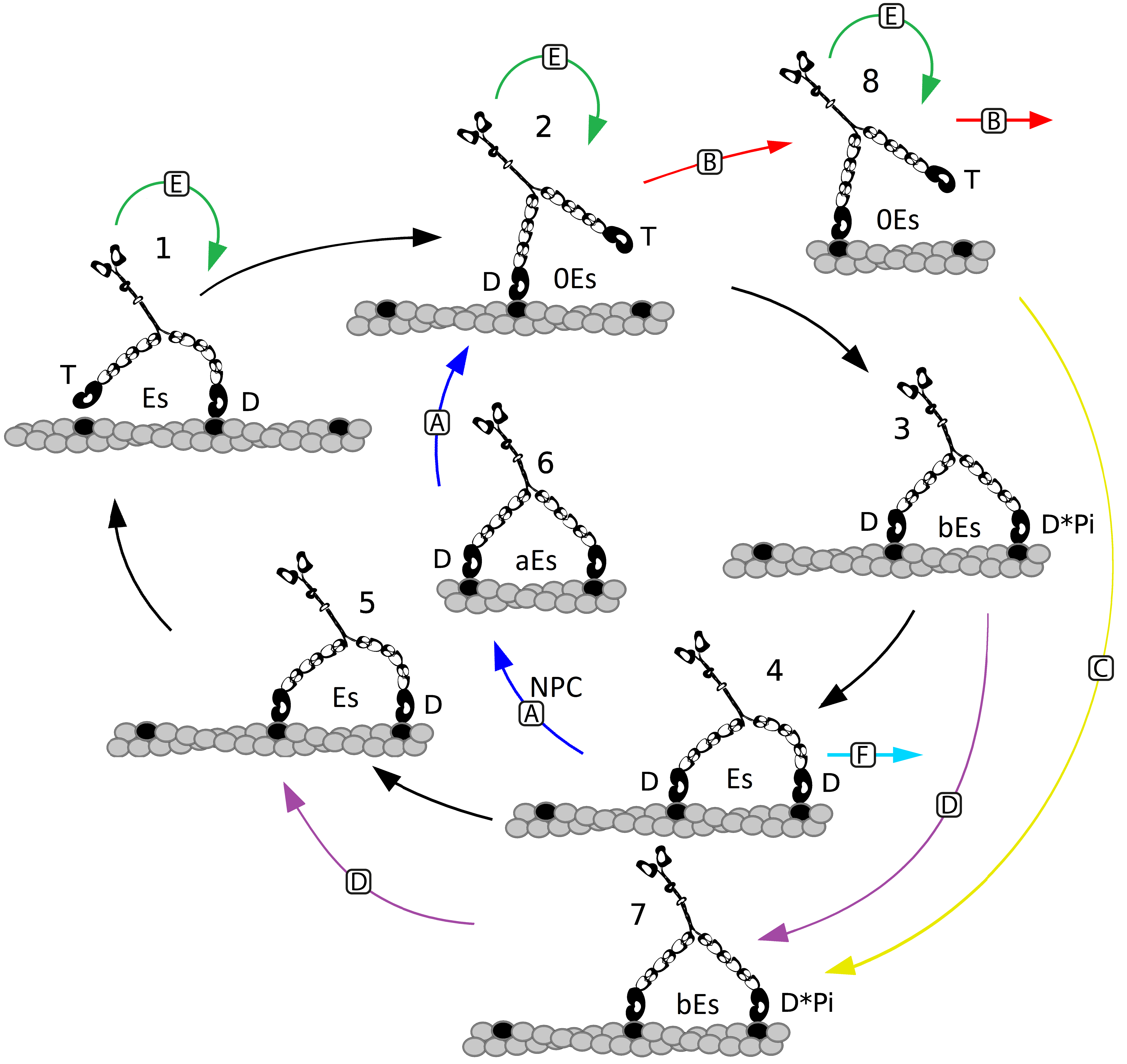}
\end{center}
\caption{
(Color online) A potential set of reaction pathways for myosin-V. Arrows denote the dominant direction of state transitions. A head labelled \textit{T} denotes a bound ATP nucleotide, \textit{D} denotes ADP and \textit{D*Pi} denotes ADP and phosphate. The internal stain energy for each state is labelled - either $E_s,~ aE_s,~bE_s$ or $0E_s$ and the nucleotide-pocket-collapse transition is labelled with NPC. Pathways: the main hydrolysis pathway (all models, black), futile cycle without (A1, blue with $a=1$) and with (A2, blue with $a<1$) NPC, chemical detachment (B, red), additional pathways (C, yellow and D, purple) and molecular slip (E, green). Mechanical detachment occurs from state 4 (F, cyan).
}
\label{fig:SAGEFM}
\end{figure}

    Our model space (Fig.~\ref{fig:SAGEFM}) comprises a set of mechanochemical states and state transitions, selected from the total set of states and transitions used in previously postulated models \cite{Baker2004,wu2007,Kad:2008wh,Bierbaum:2011dj,Skau2006,vilfan2005} and incorporating experimental evidence that suggests that ADP release is dependent upon the internal strain of the molecule \cite{rosenfeld2004,purcell2005,Oguchi:2008we,Sakamoto2008,Oke2010}. We include the following mechanisms:

    \textbf{Hydrolysis cycles} ATP is hydrolyzed at two sites within the heads of the protein producing ADP and phosphate and leading to internal strain that drives forward movement in a mechanochemically-coordinated manner \cite{veigel2002} (for example: states $4 \rightarrow 5 \rightarrow 1 \rightarrow 2 \rightarrow 3 \rightarrow 4$).

    \textbf{Futile cycle} A loss of mechanochemical coordination causing ATP hydrolysis but no forward motion ($2 \rightarrow 3 \rightarrow 4 \rightarrow 6 \rightarrow 2$). We define nucleotide pocket collapse (NPC) as a decrease in intramolecular strain as ADP is released from the front head under rearward force (transition $4 \rightarrow 6$).

    \textbf{Chemical detachment} A loss of mechanochemical coordination causing detachment from the track \cite{veigel2002,Hodges2007} ($2 \rightarrow 8 \rightarrow \ {\rm detached}$).

    \textbf{Mechanical detachment} Interaction with the bulk can cause spontaneous detachment, this is assumed to be unlikely at low external forcing (state 4 $\rightarrow {\rm detached}$ \cite{Baker2004,wu2007,Bierbaum:2011dj} ).

    \textbf{Molecular slip} Motors only weakly attached to the track can slip along it \cite{Gebhardt:2006tu,Bierbaum:2011dj} ($1 \rightarrow 1$, $2 \rightarrow 2$, $8 \rightarrow 8$).
    
    Naturally the potential model space for myosin-V is larger and can include additional mechanisms \cite{Baker2004,wu2007,Bierbaum:2011dj}, such as mechanical detachment from additional states, additional transitions and additional hydrolysis cycles. However, we present here the minimal subset that demonstrates how we use the experimental data to develop models of minimum complexity that reproduce all the observed trends of velocity and run-length with [ADP], [ATP] and external forcing.

\section{Comparison with data}

\begin{figure}
\begin{center}
\includegraphics[width=3in]{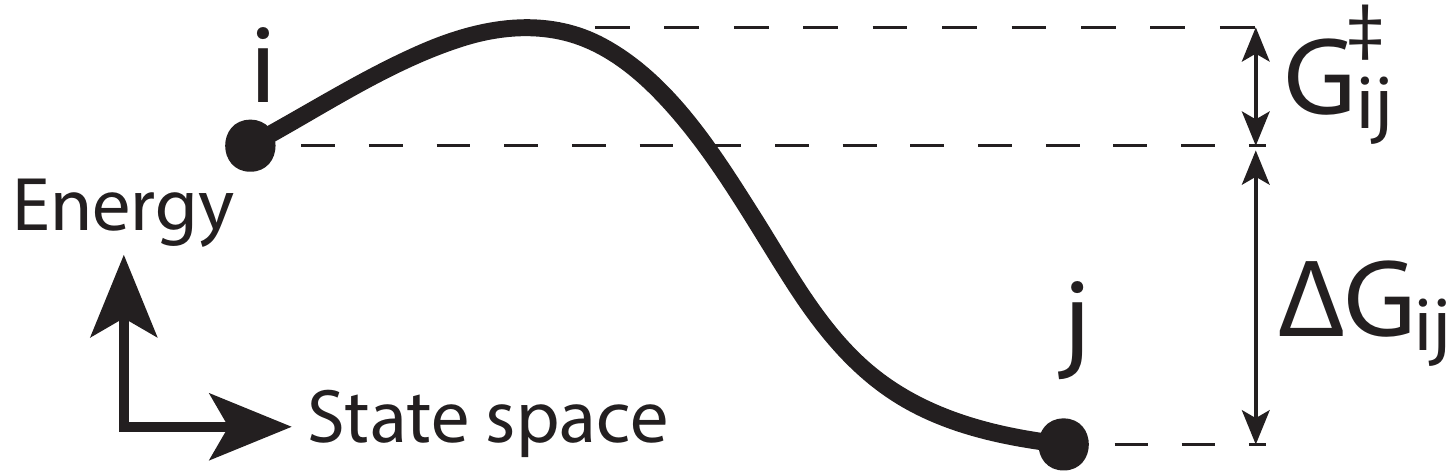}
\end{center}
\caption{A diagrammatic representation of the energy barrier $G_{ij}^{\ddag}$ and energy difference $\Delta G_{ij}$ between states $i$ and $j$ as used in Eqn.~\eqref{eqn:arrhenius}. A transition from less energetic state $j$ to more energetic state $i$ requires `climbing' $G_{ij}^{\ddag}+\Delta G_{ij}$, whereas the reverse transition only requires climbing $G_{ij}^{\ddag}$.}
\label{fig:energy}
\end{figure}

Master equations govern the state-occupancy probability dynamics and we assume a renormalized steady-state solution \cite{kolomeisky2000} (see Appendix \ref{apx:system}). Each state corresponds to a mechanochemical conformation of the molecule and the transitions between the states are described using the Arrhenius expressions
\begin{eqnarray}
    w_{ij} = \tau^{-1} \exp({-(G_{ij}^{\ddag} + \Delta G_{ij})/k_BT}),\label{eqn:arrhenius}
\end{eqnarray}
for a transition from state $i$ to state $j$ with an energy barrier $G_{ij}^{\ddag}$ and energy difference $\Delta G_{ij}$ (Fig.~\ref{fig:energy}). State transitions between any two states can take place either forwards along a cycle, in which case we denote the transition rate $u_{ij}$, or backwards, in which case we denote it $w_{ij}$ (as above). Transitions to a less energetic (usually forwards) state only involve `climbing' the energy barrier and so the energy difference term in Eqn.~\ref{eqn:arrhenius} does not appear, whereas transitions to a more energetic state (usually backwards) include both terms. Transition rates between chemically distinct states scale linearly with the relevant nucleotide concentrations. For example the forward transition from state $5$ to state $1$ in which an empty myosin-V head absorbs ATP occurs at rate
\begin{eqnarray}
    u_{51} = [ATP] \tau^{-1} \exp(-G^{\ddag}_{E-T}/k_BT).
\end{eqnarray}
Transitions where the molecule moves along the track are affected by external forcing ($f_{ex}$), i.e. the load on the motor. For example the forward transition from $2$ to $3$ where the motor takes a substep and moves a distance of $d_D$nm along its track that leads to an increase in intramolecular strain by $bE_s$ (where $b$ is a fraction and $E_s$ is the maximum strain) and ATP is hydrolyzed, occurs at rate
\begin{eqnarray}
    u_{23} = \tau_D^{-1} \exp \{-(G^{\ddag}_{T-Dw} + f_{ex} d_D + b E_s)/k_BT \}.\label{eqn:forcingExample}
\end{eqnarray}
Note that we have assumed that the distance between states in physical space is approximately the same as the distance to the corresponding energy barrier. Relaxing this assumption would be likely to improve the fit to the forcing data (Fig.~\ref{fig:results}c and f) but we do not focus upon this here. See Appendix \ref{apx:stateTrates} for a full description of the transition rates.

The velocity, dispersion and run length of the protein can be determined from the transition rates \cite{Boon:2012kt,Boon:2014}. For example the velocity $V$ is given by the forward flux through complete hydrolysis cycles and the forward slipping flux:   
\begin{eqnarray}
    	V = d\left[ \tilde{u}_{12} \tilde{P}_{1} - \tilde{w}_{21} \tilde{P}_{2} + (u_{\rm slip} - w_{\rm slip})(\tilde{P}_1 + \tilde{P}_2+\tilde{P}_8) \right], \nonumber \\
    \end{eqnarray}
    where $d=36{\rm nm}$ is the step size, $\tilde{u}_{ij}$  and $\tilde{w}_{ij}$ are the renormalized forward and backward transition rates respectively from state $i$ to state $j$ and $\tilde{P}_i$ is the renormalized steady-state state-occupancy probability.

The transition rates are defined in terms of the energetic parameters included in Eqn.~\eqref{eqn:arrhenius}. These are determined numerically using a bespoke MCSA \cite{Kirkpatrick1983} optimization routine, based on that developed by Skau et al. \cite{Skau2006}. For a given model and a given parameter set, the routine compares model predictions for dynamical quantities - such as velocities and run lengths - with a small subset of experimentally measured values and returns a cost function value $\Delta$. The parameter set that minimises $\Delta$ corresponds to the best prediction and the routine numerically explores parameter space to find this set. Optimized parameter values are subject to a sensitivity analysis. Further details are given in Appendix \ref{apx:optimisation}.

\section{Systematic model development}

A given combination of mechanisms from our model space represents a model. We aim to find a model that describes experimentally-observed relationships for the average molecular velocity ($V$) and run length ($L$) against [ATP], [ADP] and external forcing ($f_{ex}$) \cite{forkey2003,Baker2004,Kad:2008wh,Kad:2012ki,uemura2004,Gebhardt:2006tu,clemen2005} by optimizing the parameters of each model we investigate in an attempt to reproduce these trends (see Fig.~\ref{fig:results}). We found that force-dependent transition rates (as in Eqn.~\eqref{eqn:forcingExample}) and molecular slip are sufficient to give the observed experimental trends with $f_{ex}$ and so include these in all models. The particular result that has eluded all theoretical studies that are based upon rigorous physical chemistry is that $L$ decreases with both increasing [ADP] and [ATP] (denoted $L-ADP$ and $L-ATP$ respectively)\cite{Skau2006,Baker2004,wu2007,Bierbaum:2011dj}. This is where we shall focus our attention. There are at least two mechanisms that can give rise to these trends: nucleotide-dependent detachment (where molecules are more likely to leave the system as nucleotide concentration increases) or nucleotide-dependent futile cycling (where molecules are less likely to walk forwards continuously as nucleotide concentration increases) \cite{Baker2004}.

 \begin{table}[h]
\centering
\begin{tabular}{cccccccc}
\hline
\hline
\textbf{Model} & \textbf{Pathways}& \multicolumn{4}{c}{\textbf{Trends reproduced}} \\
\ & \textbf{included}& \multicolumn{2}{c}{\textbf{L vs}} & \multicolumn{2}{c}{\textbf{V vs}} \\
\ &  & [ATP] & [ADP] & [ATP] & [ADP]\\ \hline\hline
1 & D, E, F & x & \checkmark & \checkmark & \checkmark  \\
1a & B, D, E, F & x & \checkmark & \checkmark & \checkmark \\
1b & B, C, D, E, F  & \checkmark & \checkmark & \checkmark & \checkmark \\
2 & A1, B, E, F & x & \checkmark & \checkmark & \checkmark \\
2a & A1, B, C, D, E & \checkmark & x & \checkmark & \checkmark  \\
2b & A2, B, C, D, E & \checkmark & \checkmark & \checkmark & \checkmark \\
\hline
\hline
\end{tabular}
	\caption{Qualitative behavior reproduced by optimized models. \label{tab:models}}
\end{table}

\begin{figure}
\begin{center}
\includegraphics[width=2.8in]{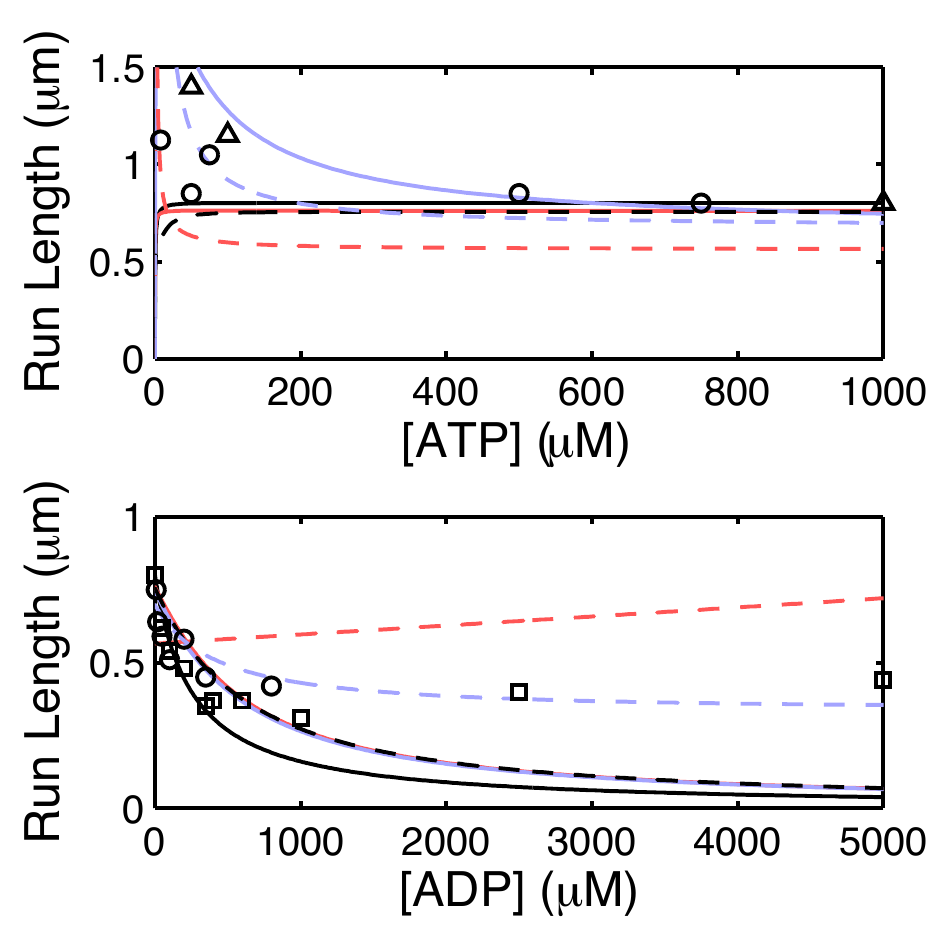}
\end{center}
\caption{ (Color online) Run length trends against [ATP] and [ADP] for model 1 (solid, black), 1a (solid, red/dark grey), 1b (solid, blue/light grey), 2 (dashed, black), 2a (dashed, red/dark grey) and 2b (dashed, blue/light grey) with [ATP]=1$\rm mM$ (lower), [Pi]=0.1$\rm \mu M$ (both) and [ADP]=0.1$\rm \mu M$ (upper). Experimentally observed relationships are shown with circles \cite{Kad:2012ki}, triangles \cite{Baker2004} and squares \cite{Baker2004}. \label{fig:Lcomparisons}}
\end{figure}

\begin{figure*}
\begin{center}
\includegraphics[width=6.1in]{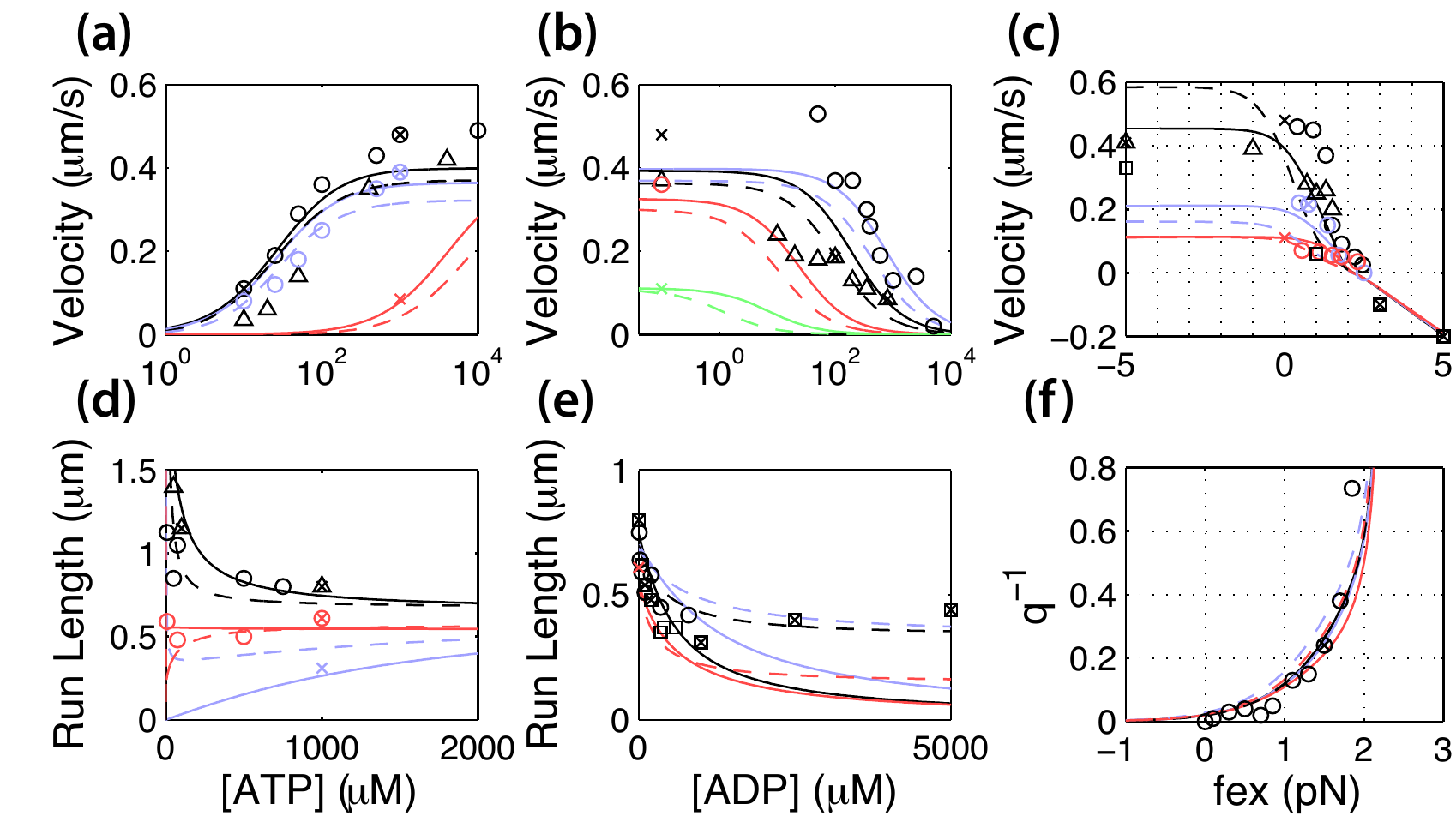}
\end{center}
\caption{ (Color online) Average velocities, run lengths and inverse step ratios as a function of [ATP], [ADP] and external forcing. Curves denote model results for models 1b (solid) and 2b (dashed), hollow shapes denote experimental data \cite{forkey2003,Baker2004,Kad:2008wh,Kad:2012ki,uemura2004,Gebhardt:2006tu,clemen2005} and crosses represent experimental data that are used in our optimization. Nucleotide concentrations are [ATP]=1$\rm mM$, [Pi]=0.1$\rm \mu M$ and [ADP]=0.1$\rm \mu M$ unless otherwise stated. Red/dark grey curves in (a) and (c)-(e) have [Pi]=40$\rm m M$.
(a) Velocity as a function of [ATP] in black, red/dark grey and blue/medium grey ([ADP]=800$\rm \mu M$) with optimization points $\Delta_{2,4,11}$ \cite{Kad:2008wh}. 
(b) Velocity as a function of [ADP] in black, red/dark grey ([ATP]=100$\rm \mu M$), blue/medium grey ([ATP]=4$\rm m M$) and green/light grey ([ATP]=10$\rm \mu M$) with optimization points $\Delta_{2,4,5,7}$ \cite{Kad:2012ki}. 
(c) Velocity as a function of $f_{ex}$ in black, red/dark grey and blue/medium grey ([ADP]=200$\rm \mu M$) with optimization points: $\Delta_{2,4,18,19,20,21}$ \cite{Kad:2012ki,uemura2004,Gebhardt:2006tu,clemen2005}.
(d) Run length as a function of [ATP] in black, red/dark grey and blue/medium grey ([ADP]=1$\rm  m M$) with optimization points $\Delta_{1,3,10}$ \cite{Baker2004,Kad:2008wh}. 
(e) Run length as a function of [ADP] in black, red/dark grey and blue/medium grey ([ATP]=2$\rm  m M$) with optimization points $\Delta_{1,6,10,12,13,14}$ \cite{Baker2004}. 
(f) The inverse step ratio as a function of $f_{ex}$ in black, red/dark grey ([ATP]=10$\rm \mu M$), blue/medium grey ([ADP]=200$\rm \mu M$) with optimization point $\Delta_{16}$ \cite{Kad:2008wh}. 
\label{fig:results}}
\end{figure*}

Model 1 is similar to that proposed by Baker et al. \cite{Baker2004}: in addition to the main hydrolysis pathway, we include an additional hydrolysis cycle, molecular slip and mechanical detachment from [ADP]-dependent state $4$ (mechanisms D, E and F, see Tab.~\ref{tab:models}). We confirm that these reproduce the observed velocity and $L-ADP$ trends (see Fig.~\ref{fig:Lcomparisons}). However there is no mechanism to give the trend for $L-ATP$. Thus we construct model 1a that adds [ATP]-dependent chemical detachment, but this is not sufficient to give $L-ATP$ because it leads to a vanishing rate of total detachment for low [ATP]. Allowing greater flexibility in the choice of hydrolysis pathway resolves this in model 1b, which reproduces all the observed experimental trends as shown in Fig.~\ref{fig:results}. 

The model proposed by Skau et al. \cite{Skau2006} includes futile cycling instead of mechanical detachment and reproduces $L-ATP$ through [ATP]-dependent chemical detachment. For low [ATP], molecules are more strongly confined to the track and so $L$ is higher. However, the model was unable to give the observed $L-ADP$ trend. We have extended the Skau model to give our model 2 by adding mechanical detachment. This gives the $L-ADP$ trend, but at low [ATP] the mechanical detachment rate is relatively large, $L$ therefore drops and so the $L-ATP$ trend is not reproduced. Model 2a includes futile cycling, chemical detachment and additional hydrolysis pathways but is unable to reproduce $L-ADP$ as failed stepping is not [ADP]-dependent. To resolve this we introduce nucleotide-dependent futile cycling. All models discussed so far assume the intra-molecular strain state in $4$ is the same as in state $6$ ($aE_s=E_s$). Relaxing this assumption is equivalent to including nucleotide pocket collapse upon ADP release; as [ADP] increases, motors become more likely to enter the futile cycle and so $L$ decreases as required for the observed $L-ADP$ trend. This is model 2b which reproduces all the experimental trends as shown in Fig.~\ref{fig:results}. 
 
A comparison of the run-length relationships is shown in Fig.~\ref{fig:Lcomparisons} and summarized in Tab.~\ref{tab:models}. Crucially only models 1b and 2b reproduce both [ATP] and [ADP] trends simultaneously. Furthermore, model 2b has non-zero run length at saturating levels of [ADP] unlike model 1b. An investigation into the sensitivity of these results to variations in the optimised parameters reveals that the qualitative results for model 2b are also more robust (as discussed in Appendix \ref{apx:optimisation}).

Each model is optimized against dynamics data as discussed, resulting in kinetic rates that correspond to specific physical processes and can be compared to measured values in the literature (Tab.~\ref{tab:modelNumbers}). The values for all of the models are reasonable to within an order of magnitude but model 2b gives the closest results for the ADP binding, ATP binding and ADP release rates. On balance this suggests the evidence is greater for a mechanism involving nucleotide-dependent futile cycling with nucleotide pocket collapse over one including mechanical motor detachment. 

 \begin{table}[h]
\centering
\begin{tabular}{ccccc}
\hline
\hline
\textbf{Source} &  \multicolumn{4}{c}{\textbf{Kinetic Rate}} \\
\ & ADP bind. & ATP bind. & Pi rel. & ADP rel. \\
\hline
Experiment  & $4$-$14$ & $0.6$-$0.9$ & $110$  & $28$-$30/0.3$-$0.4$  \\
Framework  & $w_{54}$ & $u_{51}$ & $u_{34}$ & $u_{45}$/$u_{46}$  \\
Model 1 & 1.7 & 0.44 & 110 & 13/0\\
Model 1a & 2.8 & 0.42 & 109 & 15/0\\
Model 1b & 2.9 & 0.42 & 110 & 15/0\\
Model 2 & 9.8 & 1.4 & 110 & 16/0.57\\
Model 2a & 2 & 1.7 & 110 & 14/0.44\\
Model 2b & 13.7 & 0.85 & 110 & 21/3.5\\
\hline
\hline
\end{tabular}
	\caption{Comparison of kinetic rates determined through optimization with experimental values. Units are $\mu M s^{-1}$ for the nucleotide dependent rates and $s^{-1}$ otherwise. Experimentally measured kinetic rates are shown for ADP binding\cite{DeLaCruz1999,Oke2010}, ADP release \cite{rosenfeld2004}, ATP binding \cite{DeLaCruz1999,yengo2004} and phosphate release \cite{yengo2004}. Only the rate of phosphate release is fit to in our optimization. \label{tab:modelNumbers}}
\end{table}

\section{Discussion}

We have used a guided model development process to compare candidate model architectures and hence deduce two physical-chemistry models of myosin-V stepping that are, to the best of our knowledge, the first to reproduce qualitatively all experimentally-observed velocity and run length relationships against nucleotide concentration and velocity and forward/backward step ratio trends against external forcing. 

The method we used allows us to investigate directly which aspects of highly complex models give rise to which dynamical trends, and hence to navigate intelligently through a high-dimensional model space guided by a comparison to available data. As we have demonstrated, the models we arrive at may not be unique, but the systematic comparison of reaction pathways with the experimental trends reproduced provides insight into what further data is necessary to distinguish between them.

Multiple hydrolysis pathways, molecular slip and [ATP]-dependent chemical detachment are sufficient to reproduce most of the experimental results for myosin-V stepping. However the trend of run length $L$ against [ADP] arises either from [ADP]-dependent mechanical detachment or from futile cycling that is [ADP]-dependent with the inclusion of nucleotide pocket collapse. The former reproduces the velocity against external forcing relationship more accurately but the latter is a better fit to the saturating $L$-[ADP] observations. Comparing model-predicted and experimentally observed kinetic transition rates favors the mechanism involving futile cycling and nucleotide pocket collapse. We highlight these two potential mechanisms for the walk of myosin-V for further experimental attention.

\begin{acknowledgments}
NJB was supported by the Engineering and Physical Sciences Research Council [grant number EP/P505135/1].
\end{acknowledgments}

\appendix

\section{The System}\label{apx:system}

    The state occupancy probabilities are governed by a set of master equations that describe their time evolution: 
    
      \begin{eqnarray}\numberwithin{equation}{section}\label{Skaumaster1}
        \dot{P}_1 &=& w_{21}P_2 + u_{51} P_5 - (u_{12} + w_{15} ) P_1, \\
        \dot{P}_2 &=& u_{12} P_1 + w_{32} P_3 + u_{62} P_6 + w_{82} P_8 \nonumber \\
        & &- (w_{21} + u_{23} + w_{26} + u_{28} ) P_2, \\
        \dot{P}_3 &=& u_{23} P_2 + w_{43} P_4 + w_{73} P_7 \nonumber \\
        & & - (w_{32} + u_{34} + u_{37} ) P_3, \\
        \dot{P}_4 &=& u_{34} P_3 + w_{54} P_5 + w_{64} P_6 \nonumber \\ 
        & & - (w_{43} +u_{45} + u_{46} + \delta_4  )P_4,  \\
        \dot{P}_5 &=& w_{15}P_1 + u_{45} P_4 + u_{75} P_7 \nonumber  \\
        & & - (u_{51}+w_{54} + w_{57} )P_5, \\
        \dot{P}_6 &=& w_{26} P_2 + u_{46}P_4 - (u_{62} + w_{64} )P_6,  \\
        \dot{P}_7 &=& u_{37} P_3 + w_{57} P_5 - (u_{75} + w_{78} )P_7,  \\
        \dot{P}_8 &=& u_{28}P_2 + w_{78}P_7 - (w_{82} + u_{87} + \delta_8 )P_8, \label{Skaumaster8}
      \end{eqnarray}
      where  $u_{ij}$ and $w_{ij}$ are forwards and backwards transition rates from state $i$ to state $j$ respectively. The terms $\delta_4$ and $\delta_8$ are the rates at which molecules detach from the track and are lost to the bulk owing to a mechanical and a chemical process respectively. This system can be written in matrix notation as
  \begin{equation}\numberwithin{equation}{section}
	  \dot{\mathbf{P}} = \mathbb{M} \mathbf{P}
  \end{equation}
  where $\mathbb{M}$ is a $n\times n$ reaction rate matrix and the $i^{\text{th}}$ component of vector $\mathbf{P}$ is $P_i$. Note that the equations are subject to modification for a given model (see Tab.~\ref{tab:modelConditions}).

\subsection{Renormalization}\label{sec:renormalization}

The probabilities do not sum to unity as molecules are detaching from the track. In order to use existing analytical results for motor velocity and run length \cite{Boon:2012kt,Boon:2014}, which are calculated for probability-conserving systems, we renormalise the system using the method defined by Kolomeisky and Fisher \cite{kolomeisky2000}. We write
  \begin{equation}
      P_i = \dfrac{1}{\phi_i} \exp(-\lambda t) \tilde{P}_i,
  \end{equation}
  where $\lambda$ is the dominant (closest to zero) eigenvalue of $\mathbb{M}$ and is associated with eigenvector $\phi$. Note that steady procession can only occur if detachment is linked to the slowest eigenvalue and so is slower than the other processes in the system; we assume that to be the case here. Thus we have
  \begin{equation}
   \mathbb{M}^T \mathbf{\phi} = -\lambda \mathbf{\phi},
  \end{equation}
 and so the system can now be described by
  \begin{equation}\numberwithin{equation}{section}
	  \dot{\tilde{\mathbf{P}}} = \tilde{\mathbb{M}} \tilde{\mathbf{P}}.
  \end{equation}
  $\tilde{\mathbb{M}}$ is the renormalized reaction-rate matrix with $\tilde{\delta}_4=\tilde{\delta_8}=0$ and the reaction rates
  \begin{eqnarray}
	  \tilde{u}_{ij} &=& \dfrac{\phi_j}{\phi_i} u_{ij}, \\
	  \tilde{w}_{ij} &=& \dfrac{\phi_j}{\phi_i} w_{ij}.
  \end{eqnarray}
  Dynamic quantities in our models are calculated using these renormalised rates. It can be shown \citep{renormpreprint} that the velocity of stepping motors that remain attached to actin is the same as the renormalised velocity to first order in the detachment rate. Hence the renormalised velocity can also be used to calculate the run length.

\section{State Transition Rates}\label{apx:stateTrates}

Transition rates for our models are described in Tab.~\ref{tab:rates}. The main hydrolysis cycle has $i,j \in \left[1,2,3,4,5 \right]$ and the futile cycle has $i,j \in \left[2,3,4,6 \right]$. Other hydrolysis pathways pass though states $7$ and $8$. Chemical detachment occurs from state $8$ at rate $\delta_8$ and mechanical detachment occurs from state $4$ at rate $\delta_4$. $\tau \approx 10^{-8} {\rm s}$ is the fundamental timescale of the reaction and $\tau_D \approx 10^{-5} {\rm s}$ is the hydrodynamic timescale related to movement over one step length \cite{Skau2006}. $[X]$ represents the concentration of nucleotide $X$ in the bulk. Note that some rates are modified in certain models (see Tab.~\ref{tab:modelConditions}).

\begin{center}
\begin{table}[h]
\centering
\begin{tabular}{|c|c|c|c|}
\hline
\textbf{Rates ($s^{-1}$)} & \textbf{Prefactors ($s^{-1}$)}                 & \textbf{Energies ($k_BT$)}         \\ \hline
$u_{12}$      & $\tau^{-1}_D$                           & $\theta f_{ex} d_P$                                                      \\ 
$w_{21}$      & $\tau^{-1}_D$                           & $E_s-(1-\theta)f_{ex} d_P$                                          \\ 
$u_{23}$      & $\tau^{-1}_D$                           & $G^{\ddag}_{T-Dw} + f_{ex} d_D + bE_s$                  \\ 
$w_{32}$      & $\tau^{-1}_D$                           & $G^{\ddag}_{T-Dw} + \Delta G_{T-Dw}$                   \\ 
$u_{34}$      & $\tau^{-1}$                             & $G^{\ddag}_{Dw-Ds}$                                    \\ 
$w_{43}$      & $\rm{[ Pi ]} \tau^{-1}$                & $G^{\ddag}_{Dw-Ds}+\Delta G_{Dw-Ds}$  \\ 
              &  & $- (1-b) E_s - f_{ex} d_B$ \\
$u_{45}$      & $\tau^{-1}$                             & $G^{\ddag}_{Ds-E}$                                     \\ 
$w_{54}$      & $\rm{[ ADP ]} \tau^{-1}$                & $G^{\ddag}_{Ds-E} + \Delta G_{Ds-E}$                   \\ 
$u_{51}$      & $\rm{[ ATP ]} \tau^{-1}$                & $G^{\ddag}_{E-T}$                                      \\ 
$w_{15}$      & $\tau^{-1}$                             & $G^{\ddag}_{E-T} + \Delta G_{E-T}$                     \\
$u_{37}$      & $u_{45}$                                & $\beta E_s$ \\
$w_{73}$      & $w_{54}$                                & $\beta E_s$ \\ 
$u_{46}$      & $u_{45}$                     & $\alpha E_s$ \\
$w_{64}$      & $w_{54}$                     & $(1-a)E_s + \alpha E_s$ \\
$u_{62}$      & $u_{51}$                     & $\alpha E_s$ \\
$w_{26}$      & $w_{15}$                     & $aE_s + \alpha E_s + f_{ex}(d_D + d_B)$\\ 
$u_{28}$      & $u_{45}$                     & $\gamma E_s$ \\ 
$w_{82}$      & $w_{54}$                     & $\gamma E_s$ \\ 
$u_{87}$      & $u_{23}$                     & 0 \\ 
$w_{78}$      & $w_{32}$                     & 0 \\ 
$\delta_8$    & $u_{51}$                     & $-|f_{ex}|\epsilon$ \\
$\delta_{4}$  & $\tau^{-1}$                     & $G^{\ddag}_{\delta 4}$ \\ 
\hline
\end{tabular}\label{tab:rates}
	\caption{An energetic description of the transition rates within the models. If $W_{ij}$ is the rate from state $i$ to $j$, $p$ is the prefactor and $E_{ij}$ is the energy required for this transition, then $W_{ij} = p e^{-E_{ij}/k_BT}$. Parameters are described in the text; those to be optimized are listed in Tab. \ref{tab:optpar}}.
\end{table}
\end{center}

\begin{center}
 \begin{table}[h]
\centering
\begin{tabular}{ccc}
\hline
\hline
\textbf{Model} & \textbf{Pathways}& \textbf{Conditions}  \\ 
\hline
1 & D, E, F & $a=1$, $\alpha E_s = 0$, $\gamma E_s = 0$, \\
& & $u_{46}=w_{64}=u_{28}=w_{78}=0$ \\
1a & B, D, E, F & $a=1$, $\alpha E_s = 0$, \\
& & $u_{46}=w_{64}=u_{37}=w_{57}=u_{87}=0$  \\
1b & B, C, D, E, F  & $a=1$, $\alpha E_s = 0$, $u_{46}=w_{64}=0$  \\
2 & A1, B, E, F & $a=1$, $\beta E_s = 0$, \\
& & $u_{37}=w_{57}=u_{87}=0$ \\
2a & A1, B, C, D, E & $a=1$, $\delta_4=0$ \\
2b & A2, B, C, D, E & $\delta_4=0$ \\
\hline
\end{tabular}
	\caption{Conditions on model parameters implied by the inclusion of the particular pathways or mechanisms in each model. Pathways are defined in Fig. 1 of the main text, transition rates are defined in equations (\ref{Skaumaster1})-(\ref{Skaumaster8}), and the remaining parameters are defined in the text and in Tab. \ref{tab:rates}. \label{tab:modelConditions}}
\end{table}  
\end{center}

In this study there are four distinct chemical energy differences relating to the changing chemical states of the myosin heads. Moving from a state in which the head is not attached to the track and has a bound ATP nucleotide and to the state in which the head is attached and has ADP and P$_i$ nucleotides bound is associated with an energy difference $\Delta G_{T-Dw}$. Transitioning from this to a state with only ADP bound corresponds to an energy difference of $\Delta G_{Dw-Ds}$. The subsequent release of the ADP-bound nucleotide gives an energy difference of $\Delta G_{Ds-E}$ and then detachment from the track and binding of a ATP nucleotide to the myosin head leads to an energy difference $\Delta G_{E-T}$. Similar notation, $G^{\ddag}_{T-Dw}$, $G^{\ddag}_{Dw-Ds}$, $G^{\ddag}_{Ds-E}$, $G^{\ddag}_{E-T}$, is used to describe the chemical energy barriers between states. 

There are several mechanical energy differences: changes in the internal strain of the motor and energies relating to movement along the track. States $1$, $4$ and $5$ are maximally strained, with internal strain equal to $E_s$, and states $2$ and $8$ are unstrained. There are two intermediate levels of strain: $bE_s$ in states $3$ and $7$ and $a E_s$ in state $6$. The main powerstroke step (transition $1 \rightarrow 2$) is modelled as a complete release of $E_s$. A subsequent small diffusive step (transitions $2 \rightarrow 3$ or $8 \rightarrow 7$) corresponds to a small increase in internal strain $0 \rightarrow b E_s$. Strong binding of the front myosin head to the actin induces the internal strain increase $b E_s \rightarrow E_s$ (transitions $3 \rightarrow 4$ or $7 \rightarrow 5$). 

Capello et al. \cite{cappello2007} observe three steps in the walk of myosin-V of $d_P \approx 23$nm, $d_D \approx 8$nm and $d_B \approx 5$nm. In our models, movement over the distance $d_P$ corresponds to a complete release of the maximum internal strain energy $E_s \rightarrow 0$, $d_D$ corresponds to an increase from no internal strain to a partially strained state $0 \rightarrow bE_s$ and $d_B$ corresponds to a further increase to maximum strain $bE_s \rightarrow E_s$. Assuming that when only one head is attached to actin the molecule behaves as a Hookean spring we have
\begin{equation}
    \Delta E_s = \dfrac{1}{2} k_H \Delta d^2.
\end{equation}
Thus $b = d_D^2 / d_P^2$ with $d = d_P + d_D + d_B = 36$nm \cite{Mehta1999}. Step sizes were chosen from the literature \cite{cappello2007}, $d_P=23$nm, $d_D = 8$nm and $d_B = 5$nm. Motion over these distances requires energy of $f_{ex}d_P$, $f_{ex}d_D$ and $f_{ex}d_B$ respectively, where $f_{ex}$ is the component of the pico-newton size external force parallel to the direction of motion of the motor owing to the motor's cargo. We introduce an additional parameter $\theta$, a load distribution factor \cite{Bierbaum:2011dj,Fisher1999,Fisher:1999wz,Kolomeisky2003} to tune the interaction of the main powerstroke step (transition $1 \rightarrow 2$) with $f_{ex}$.

Following Skau et al. \cite{Skau2006} we define $\Gamma$ to be a measurement of the deviation of the system from equilibrium
    \begin{eqnarray}\numberwithin{equation}{section}
        \Gamma = \prod_{i \in \rm cycle} \dfrac{u_i}{w_i} = e^{(\Delta G_{\rm hyd} - f_{ex} d) / k_B T}.
    \end{eqnarray}
$\Delta G_{\rm hyd}$ is the total energy difference for the ATP hydrolysis and is calculated to be approximately $25 k_BT$ at cellular conditions with the standard free energy being approximately $13 k_BT$ \cite{Alberty1992}. At equilibrium we have $\Gamma=1$ to fulfil detailed balance \cite{Onsager:1931uu,Liepelt:2007hs,Lipowsky:2007db} and this gives a thermodynamic upper bound on the stall force $f_{\rm stall} = \Delta G_{\rm hyd} / d \approx 2.8pN$.

In our models there are two mechanisms through which myosin-V can detach from the track: chemical or mechanical. The chemical detachment occurs when ATP binds to the only attached myosin head in state $8$ at rate $\delta_8$ and so corresponds to a loss of coordination between the heads. Mechanical detachment describes the molecule being physically pulled or knocked off the track from state $4$ and is assumed to occur at a constant rate $\delta_4$. External forcing increases the probability of chemical detachment and thus \cite{Skau2006}
\begin{eqnarray}
 \delta_8(f_{ex}) = \delta_8(0) \exp(|f_{ex}|\epsilon/k_BT),
\end{eqnarray}
where $\epsilon=2.4{\rm nm}$ is the interaction distance approximated from single myosin head pulling experiments \cite{nishizaka2000}.
  
    An additional pathway that may be involved at high external forcing - a jump from one cycle repeat to another - is adapted from Bierbaum et al. \cite{Bierbaum:2011dj}:
    \begin{eqnarray}
    	w_{\rm slip} &=& \dfrac{D'(f_{ex} d - U^{\ddag})}{d^2 k_BT}(1- e^{(U^{\ddag}-f_{ex} d)/k_BT})^{-1}, \\
    	u_{\rm slip} &=& w_{\rm slip} e^{-f_{ex} d/k_BT},
    \end{eqnarray}
    where $D'=300$ nm$^2/$s is the diffusion constant. Here we depart from the value chosen by Bierbaum et al. \cite{Bierbaum:2011dj} as our own analysis suggests this lower value of $D'$ gives a better fit to the authors' results. $U^{\ddag}$ is the energy barrier for slipping; the authors chose $U^{\ddag}=20 k_BT$ in their study and this is what we use here. Physically these transitions correspond to a postulated forwards and backwards slipping respectively from one cycle to the next. It has been shown that reversed motion down the track is independent of ATP \cite{Gebhardt:2006tu}, and so a slipping process that is nucleotide independent accords with current knowledge. In these models it is assumed that slipping can only happen from states in which only one myosin head is bound to the track (states $1$, $2$ and $8$) to the same state. Therefore $w_{11} = w_{22} = w_{88} = w_{\rm slip}$ and $u_{11} = u_{22} = u_{88} = u_{\rm slip}$. These rates have no effect on the governing state-space master equations but do have an influence on the velocity as each molecule that undergoes such a transition slips $36$nm along the actin filament.
    
    Using the renormalization method (Sec.~\ref{sec:renormalization}), the detachment rates are set to zero and the probabilities, and transition rates are scaled appropriately resulting in a probability-conserving system. The motor velocity $V$ and dispersion $D$ can therefore be determined analytically by methods described by Boon and Hoyle \cite{Boon:2012kt,Boon:2014}. The velocity of molecules is
    \begin{eqnarray}
    	V = d\left[ \tilde{u}_{12} \tilde{P}_{1} - \tilde{w}_{21} \tilde{P}_{2} + (u_{\rm slip} - w_{\rm slip})(\tilde{P}_1 + \tilde{P}_2 + \tilde{P}_8) \right]. \nonumber \\
    \end{eqnarray}
Note that only transitions from one repeat of a stepping cycle to another need including in the above expression. The run length \cite{kolomeisky2000} is $L = V / \lambda$, where $\lambda$ is the dominant eigenvalue of the transposed reaction rate matrix $\mathbb{M}^T$. The forwards/backwards step ratio is given by
    \begin{eqnarray}
    	q = \dfrac{ \tilde{u}_{12} \tilde{P}_1 + u_{\rm slip} (\tilde{P}_1 + \tilde{P}_2 + \tilde{P}_8)}{ \tilde{w}_{21} \tilde{P}_2 + w_{\rm slip} (\tilde{P}_1 + \tilde{P}_2 + \tilde{P}_8)}.
    \end{eqnarray}

\section{The Optimization}\label{apx:optimisation}

Skau et al. \cite{Skau2006} constructed an optimization procedure to fit a discrete stochastic model for the myosin-V stepping cycle to experimental data. We have modified their method to determine the best-fit parameters for each model architecture. See Supplemental Material at [URL will be inserted by publisher] for a collection of the core MATLAB routines that we created for our optimization.
    
Transition rates are calculated from a choice of energetic parameters. The validity of a given set of these is determined numerically by the degree to which the model results match experimental data; we fit a given model to energetic, velocity, run length and forwards/backwards step ratio data under cellular conditions \cite{Baker2004,DeLaCruz1999,DeLaCruz2000,wang2000,rief2000,mehta2001,howard2001,forkey2003,yildiz2003,uemura2004,Kad:2008wh}. Unlike other studies \cite{Baker2004,Kad:2012ki,wu2007,Bierbaum:2011dj} we do not fit transition rates to kinetic data (with the exception of the phosphate release rate). Thus these values are not a priori expected to match experimental kinetic data. We extract the model kinetic rates from the dynamics data that we fit to. The degree of agreement of the kinetic rates with experiment is evidence to support the validity of a model. In addition, we do not provide every experimentally-observed data point to the optimization; instead, we specify a limited set (labelled $\Delta_i$ with $i \in [1,...,27]$, marked where possible on the figures in the main text by crosses) and observe whether a given model architecture can reproduce the qualitative behavior of data defined by the remaining points (not used in the optimisation).

In our models there are effectively up to 13 free parameters that we optimize against experimental data: the chemical energy differences $\Delta G_{T-Dw}$, $\Delta G_{Dw-Ds}$, $\Delta G_{Ds-E}$ and $\Delta G_{E-T}$, whose values are known approximately (encoded by four additional terms in the optimization) and fixed to sum to approximately $13k_BT$ \cite{Skau2006}; the chemical energy barriers defined by $G^{\ddag}_{T-Dw}$, $G^{\ddag}_{Dw-Ds}$, $G^{\ddag}_{Ds-E}$ and $G^{\ddag}_{E-T}$; the internal molecular strain values set by $E_s$ and $aE_s$ with $E_s>aE_s$; the ADP-gating energy barriers $\alpha E_s$, $\beta E_s$ and $\gamma E_s$; $G^{\ddag}_{\delta 4}$ that determines the constant rate of detachment from state $4$; and the load distribution factor $\theta$ \cite{Kolomeisky2003} that tunes the interaction of the main powerstroke step with $f_{ex}$.

Our bespoke simulated-annealing \cite{Kirkpatrick1983} optimization routine explores parameter space to find the combination of parameters that enables a given model to reproduce experimental results most accurately. This is measured by the cost function $\Delta$. The extensive exploration of a high-dimensional parameter space to find starting points for our routine is numerically expensive: to improve computational efficiency we estimate the starting point based on established results. For the free parameters included in the Skau model \cite{Skau2006} we use the optimized values found by Skau et al. The initial values of the additional parameters are chosen to be $aE_s=E_s$, $\beta E_s = \gamma E_s = 0$ and $\theta=0$, again to match the Skau model.
    
    In addition to the Skau initial point, $10,000$ random start points were also selected and optimized from in order to check for other low cost regions. Each run with a low cost result moved back towards the region in which the Skau parameters are located. Those that were not low cost became stuck in high-cost local energy wells. Once a low cost point for a particular model was identified, we performed an analysis of the surrounding hypersurface in parameter space in order to assess the robustness of the model at that point.

\subsection{Cost Function}\label{apx:cost}

The cost function contains $27$ terms
    \begin{eqnarray}
	    \Delta = \sum_{i=1}^{27} \Delta_i(\rm [ATP],[ADP],[Pi],f_{ex}),
    \end{eqnarray}
and each compares a result from a model against experimental data using a least-squares method
    \begin{eqnarray}
	   \Delta_i(\rm{[ATP],[ADP],[Pi],f_{ex}}) = \dfrac{(R-R_E)^2}{\sigma^2_{R_E}},
    \end{eqnarray}
where $R$ is the model result, $R_E$ is the experimental result and $\sigma^2_{R_E}$ is the mean-squared uncertainty in the experimental result; each is dependent on conditions $\rm [ATP]$, $\rm [ADP]$, $\rm [Pi]$ and $\rm f_{ex}$.

\begin{widetext}
\begin{center}
   \begin{table}
	\centering
	\begin{tabular}{ | c | c | c | c | c | }
    	\hline
    	$\Delta_i(\rm [ATP],{\rm [ADP]},{\rm [Pi]},f_{\rm ex})$ & $R$ & $R_E$ & $\sigma_{R_E}$ & Studies \\
    	\hline
    	$\Delta_1(1\rm{mM}, 0.1\rm{\mu M}, 0.1\rm{\mu M},0)$ & $L$ & $0.8 \rm{\mu m}$ & $0.015$ & \cite{Baker2004} \\
    	$\Delta_2(1\rm{mM}, 0.1\rm{\mu M}, 0.1\rm{\mu M},0)$ & $V$ & $0.48 \rm{\mu m}s^{-1}$ & $0.02$ & \cite{Kad:2012ki} \\
    	$\Delta_3(100 \rm{\mu M}, 0.1\rm{\mu M}, 0.1\rm{\mu M},0)$ & $L$ & $1.15 \rm{\mu m}$ & $0.15$ & \cite{Baker2004} \\
    	$\Delta_4(10 \rm{\mu M}, 0.1\rm{\mu M}, 0.1\rm{\mu M},0)$ & $V$ & $0.11 \rm{\mu m}s^{-1}$ & $0.011$ & \cite{Kad:2012ki} \\
    	$\Delta_5(1\rm{mM}, 100\rm{\mu M}, 0.1\rm{\mu M},0)$ & $V$ & $0.185 \rm{\mu m}s^{-1}$ & $0.0185$ & \cite{Kad:2012ki} \\
    	$\Delta_6(1\rm{mM}, 2.5\rm{mM}, 0.1\rm{\mu M},0)$ & $L$ & $0.4 \rm{\mu m}$ & $0.15$ & \cite{Baker2004} \\
    	$\Delta_7(1\rm{mM}, 800\rm{\mu M}, 0.1\rm{\mu M},0)$ & $V$ & $0.085 \rm{\mu m}s^{-1}$ & $0.0085$ & \cite{Kad:2012ki} \\
    	$\Delta_8(1\rm{mM}, 0.1 \rm{\mu M}, 4\rm{mM}, 0)$ & $L$ & $0.5 \rm{\mu m}$ & $0.15$ & \cite{Baker2004,mehta2001} \\
    	$\Delta_9(1\rm{mM}, 0.1 \rm{\mu M}, 4\rm{mM}, 0)$ & $V$ & $0.44 \rm{\mu m}s^{-1}$ & $0.044$ & \cite{Baker2004,mehta2001} \\
    	$\Delta_{10}(1\rm{mM}, 0.1 \rm{\mu M}, 40\rm{m M}, 0)$ & $L$ & $ 0.61\rm{\mu m}$ & $0.061$ & \cite{Kad:2012ki} \\
    	$\Delta_{11}(1\rm{mM}, 0.1 \rm{\mu M}, 40\rm{m M}, 0)$ & $V$ & $ 0.39\rm{\mu m}s^{-1}$ & $0.039$ & \cite{Kad:2012ki} \\
    	$\Delta_{12}(1\rm{mM}, 200 \rm{\mu M}, 0.1\rm{m M}, 0)$ & $L$ & $0.48 \rm{\mu m}$ & $0.05$ & \cite{Baker2004}  \\
    	$\Delta_{13}(1\rm{mM}, 1 \rm{m M}, 0.1\rm{m M}, 0)$ & $L$ & $0.31 \rm{\mu m}$ & $0.025$ & \cite{Baker2004}  \\
    	$\Delta_{14}(1\rm{mM}, 5 \rm{m M}, 0.1\rm{m M}, 0)$ & $L$ & $0.44 \rm{\mu m}$ & $0.05$ & \cite{Baker2004}  \\
    	$\Delta_{15}(1\rm{mM}, 100 \rm{\mu M}, 0.1\rm{m M}, 0)$ & $L$ & $0.54 \rm{\mu m}$ & $0.05$ & \cite{Baker2004}  \\
    	$\Delta_{16}(100 \rm{\mu M}, 0.1 \rm{\mu M}, 0.1\rm{m M}, 1.5)$ & $q$ & $0.24 $ & $0.01$ & \cite{Kad:2008wh}  \\
    	$\Delta_{17}(1\rm{mM}, 200\rm{\mu M}, 0.1\rm{\mu M},0.75 pN)$ & $L$ & $0.4 \rm{\mu m}$ & $0.15$ & \cite{uemura2004} \\
    	$\Delta_{18}(1\rm{mM}, 200\rm{\mu M}, 0.1\rm{\mu M},0.75 pN)$ & $V$ & $0.215 \rm{\mu m}s^{-1}$ & $0.05$ & \cite{uemura2004} \\
    	$\Delta_{19}(1\rm{\mu M}, 0.1 \rm{\mu M}, 0.1\rm{\mu M},5 pN)$ & $V$ & $-0.2 \rm{\mu m}s^{-1}$ & $0.05$ & \cite{Gebhardt:2006tu} \\
    	$\Delta_{20}(1\rm{\mu M}, 0.1 \rm{\mu M}, 0.1\rm{\mu M},3 pN)$ & $V$ & $-0.1 \rm{\mu m}s^{-1}$ & $0.05$ & \cite{Gebhardt:2006tu} \\
    	$\Delta_{21}(2\rm{\mu M}, 0.1 \rm{\mu M}, 0.1\rm{\mu M},-5 pN)$ & $V$ & $0.41 \rm{\mu m}s^{-1}$ & $0.025$ & \cite{clemen2005} \\
    	\hline
  	\end{tabular}
	\caption{The experimentally measured data points for the run length, molecular velocity and step ratio included in the cost function }\label{tab:CostFunctionPoints}
\end{table}
\end{center}
\end{widetext}

All cost function terms pertaining to dynamic quantities are listed in Tab.~\ref{tab:CostFunctionPoints}. The velocity and run length of myosin-V have been measured experimentally under varying nucleotide concentrations \cite{Baker2004,Kad:2012ki,DeLaCruz1999,DeLaCruz2000,wang2000,rief2000,forkey2003,yildiz2003,uemura2004} and terms $1$-$15$ in the cost function represent these measurements. Term $16$ represents the measured forwards/backwards step ratio \cite{Kad:2008wh}. Terms $17$-$21$ are based on the velocity and the run length measurements of myosin-V molecules under external forcing \cite{Gebhardt:2006tu,uemura2004,clemen2005}.

The next four terms of the cost function represent energetic restrictions on the interaction of the protein with the actin track \cite{howard2001}:
    \begin{eqnarray}
	    \Delta_{22} &=& \left(\dfrac{\Delta G_{T-Dw}-2 k_BT}{3 k_BT}\right)^2, \\
	    \Delta_{23} &=& \left(\dfrac{\Delta G_{Dw-Ds}-5.7 k_BT}{3 k_BT}\right)^2, \\
	    \Delta_{24} &=& \left(\dfrac{\Delta G_{Ds-E}+7.7 k_BT}{3 k_BT}\right)^2, \\
	    \Delta_{25} &=& \left(\dfrac{\Delta G_{E-T}-15.3 k_BT}{3 k_BT}\right)^2.
    \end{eqnarray}
The reaction energy differences are restricted so that they sum to the standard free energy in one hydrolysis cycle \cite{Skau2006}: 
	\begin{eqnarray}
		& &\Delta G_{T-Dw}+\Delta G_{Dw-Ds}+\Delta G_{Ds-E}+\Delta G_{E-T} \nonumber \\
		& & = 13.125 k_B T. 
	\end{eqnarray}
	
The next term in the cost function ensures that the release of ADP from the front head is much slower than that from the rear
    \begin{eqnarray}
	    \Delta_{26} &=& \Delta^{\rm max}_{26} (u_{46} / u_{45}),
    \end{eqnarray}
as shown by experiment \cite{rosenfeld2004,purcell2005,Oguchi:2008we,Sakamoto2008}. Note that as $u_{45} \ge u_{46}$ we can choose $\Delta^{\rm max}_{26}=50$ to weight this optimization point sufficiently relative to the others.

We found that terms 1-26 in the cost function fail to fix the phosphate release rate sufficiently. Thus the last term in the cost function does exactly this using data from Yengo et al. \cite{yengo2004}:
    \begin{eqnarray}
	    \Delta_{27} &=& \left(u_{34}-110\right)^2.
    \end{eqnarray}

27 terms in the cost function and only 12-13 effective parameters to fit is evidence to suggest that a low-cost solution is unlikely to be found unless the model architecture can give the experimental results naturally and without curve fitting.

\subsection{Optimized Parameters}\label{apx:sensitivity}

\begin{figure*}
\begin{center}
\includegraphics[width=6.35in]{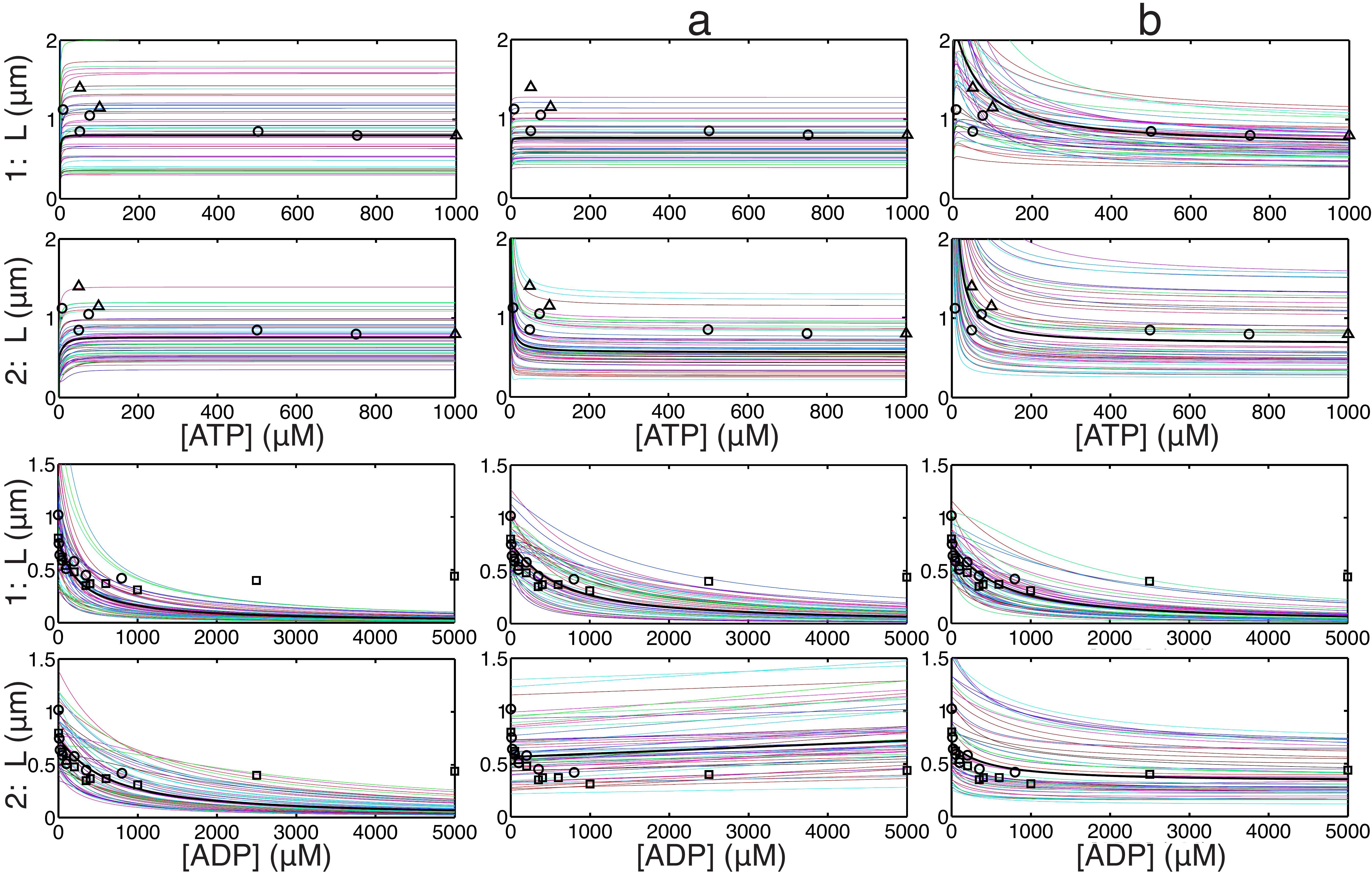}
\end{center}
\caption{ (Color online) Investigation of the sensitivity of the run length behavior: $50$ randomly generated parameter sets where all optimised parameters are varied by up to $5$\% from the nominal values. Run length curves are shown for models 1-1b (first and third rows) and 2-2b (second and last rows). \label{fig:sensitivity}}
\end{figure*}

The optimized parameter values for each model we investigated are listed in Table \ref{tab:optpar}. We also perform an investigation into the sensitivity of the results presented in Tab.~\ref{tab:models} to variations in the optimised parameters. Fig.~\ref{fig:sensitivity} demonstrate these results for the run length. We find that the qualitative behavior of the run length in model 2b is more robust to parameter variation than in model 1b. 

\begin{widetext}
\begin{center}
\begin{table}[!ht]
\centering
	\begin{tabular}{ |c|c|c|c|c|c|c| }
    	\hline
	  Parameter & Model 1 & Model 1a & Model 1b & Model 2 & Model 2a & Model 2b \\
	  \hline
	  $\Delta G_{T-Dw}$     & 2.46    & 2.07 & 1.78 & 1.67 & 2.10 & 1.63 \\
	  $\Delta G_{Dw-Ds}$    & 8.33    & 8.15 & 8.91 & 8.82 & 10.05& 9.26 \\
	  $\Delta G_{Ds-E}$     & -11.81  &-12.17&-12.18&-13.35&-11.85&-13.39 \\
	  $\Delta G_{E-T}$      & 14.14   & 15.08& 14.61& 15.98& 12.82& 15.62 \\
	  $G^{\ddag}_{T-Dw}$    & 4.61    & 5.11 & 5.13 & 5.36 & 7.47 & 6.37 \\
	  $G^{\ddag}_{Dw-Ds}$   & 13.72   & 13.73& 13.73& 13.72& 13.72& 13.72 \\
	  $G^{\ddag}_{Ds-E}$    & 15.86   & 15.73& 15.72& 15.67& 15.75& 15.38 \\
	  $G^{\ddag}_{E-T}$     & 5.44    & 5.47 & 5.46 & 4.30 & 4.09 & 4.77 \\
	  $G^{\ddag}_{\delta 4}$& 18.96   & 19.89& 19.89& 19.93& n/a  & n/a\\
	  $E_{s}$               & 10.51   & 10.81& 10.92& 11.24& 13.07& 12.15 \\
	  $a E_s$               & n/a     & n/a  & n/a  & n/a  & n/a  & 4.38 \\
	  $\alpha E_s$          & n/a     & n/a  & n/a  & 3.31 & 3.50 & 1.80 \\
	  $\beta E_s$           & 6.58    & 7.82 & 7.69 & n/a  & 6.51 & 6.78 \\
	  $\gamma E_s$          & n/a     & 1.56 & 1.51 & 1.97 & 3.04 & 2.70 \\
	  $\theta$              & 1.00    & 1.00 & 1.00 & 1.00 & 1.00 & 1.00 \\
    	\hline
   	\end{tabular}
\caption{\label{tab:optpar} Optimized parameters for our models. The first four chemical energy differences are fixed to sum to approximately $13k_BT$. The next four pertain to chemical energy barriers, while $G^{\ddag}_{\delta 4}$ is the energy barrier for mechanical detachment, $E_s$/$aE_s$ are the maximum/intermediate levels of internal intramolecular strain respectively, $\alpha E_s$, $\beta E_s$ and $\gamma E_s$ are ADP release gating energy barriers and $\theta$ mediates the interaction of the powerstroke with $f_{ex}$. The units for all the parameters is $k_BT$, except $\theta$ which is dimensionless and lies between 0 and 1.}
\end{table}
\end{center}
\end{widetext}

\end{document}